\documentclass[aip,jcp,superscriptaddress,groupedaddress,11pt]{revtex4}  
\usepackage{amsfonts}
\usepackage{amsmath}
\usepackage{amssymb}
\usepackage{version}
\usepackage{graphicx}%
\includeversion{New_connection}
\excludeversion{Old_connection}

\begin{document}
\title{The Effects of Intrinsic Dynamical Ghost Modes in Discrete-Time Langevin Simulations}
\author{Lucas Frese Gr{\o}nbech Jensen}
\thanks{Current Address: Department of Applied Mathematics and Computer Science, The Technical University of Denmark, 2800 Kgs.~Lyngby, Denmark}
\affiliation{Department of Mathematics, University of California, Santa Barbara, CA 93106, USA}
\author{Niels Gr{\o}nbech-Jensen}
\email{ngjensen@ucdavis.edu}
\affiliation{Department of Mechanical and Aerospace Engineering, University of California, Davis, CA 95616, USA}
\affiliation{Department of Mathematics, University of California, Davis, CA 95616, USA}
\begin{abstract}
Using the recently published GJF-2GJ Langevin thermostat, which can produce time-step-independent statistical measures even for large time steps, we analyze and discuss the causes for abrupt deviations in statistical data as the time step is increased for some simulations of nonlinear oscillators. Exemplified by the pendulum, we identify a couple of discrete-time dynamical modes in the purely damped pendulum equation as the cause of the observed discrepancies in statistics. The existence, stability and kinetics of the modes are consistent with the acquired velocity distribution functions from Langevin simulations, and we conclude that the simulation deviations from physical expectations are not due to normal, systematic algorithmic time-step errors, but instead due to the inherent properties of discrete time in nonlinear dynamics.
\end{abstract}
\maketitle
February 3, 2019
\section{Introduction}
\label{sec:intro}
For decades, discrete-time Langevin dynamics has been a subject of considerable importance in computational
statistical mechanics, which relies on two main tools; namely Monte Carlo sampling or Langevin dynamics (see, e.g., Refs.~\cite{Frenkel,Hoover}). The former has the advantage of being guaranteed to sample the correct Boltzmann distribution if enough time is offered to the task, however, the efficiency of sampling is not always obvious, and the method is limited in its ability to provide temporal information. The latter offers the ability to mimic time evolution, and the sampling strategy is given by the simultaneous mutual interactions in the entire system, however, the discrete-time evolution is known to introduce systematic errors as the time step is increased, as seen in, e.g., the well-known stochastic thermostats in Refs.~\cite{SS,BBK,Pastor_88}. Recent developments have offered the possibility of conducting discrete-time stochastic Langevin simulations with seemingly no (or at least very small) errors in the obtained statistics of both configurational \cite{ML,GJF1} and simultaneous kinetic sampling \cite{2GJ}. The time-step-independent  sampling is analytically proven for linear systems (linear and harmonic potentials), and the applications to nonlinear and complex systems, such as condensed molecular ensembles, have been demonstrated with very good results \cite{GJF1,GJF2,GJF4,2GJ}, especially for molecular simulations of condensed phases. As a result, it is tempting to challenge the time step to its limit with the expectation that any system will respond according to the linear analysis. In doing that, some simulations have shown curious deviations from these attractive features for large time-steps \cite{GJF4}, especially for systems with relatively low dissipation. The observed deviations are not of the typical kind that indicate systematic deviations which are usually associated with discrete-time algorithms \cite{Pastor_88}. These systematic deviations can be evaluated by linear analysis, and linear analysis of the recent method in Ref.~\cite{2GJ} has no systematic errors in either configurational or kinetic sampling. Instead, the deviations under investigation here seem to be absent for small time steps, and then abruptly appear at a certain threshold below the stability limit of the method. This is reminiscent of previous observations of nonlinear resonant artifacts in Molecular Dynamics simulations \cite{Schlick_98,Ma}, which have become a field of intense study in Molecular Dynamics \cite{LMandMT}. We will explore this phenomenon in the framework of the very simple, damped, noisy pendulum equation in order to illuminate the issue from a nonlinear dynamics viewpoint. Such corrugated potential, in which transport and diffusion are subjects of continued interest, periodic or tilted, is in of itself a system of both fundamental interest and relevance for many applications in condensed matter physics, e.g., Josephson applications \cite{VW,Kautz,Blackburn_book,JJ_review} or atomic surface diffusion \cite{surface1}, the latter having been further numerically addressed in Ref.~\cite{Voter} in order to statistically enhance simulated barrier transitions.

The equation of interest is the Langevin equation
\cite{Langevin}
\begin{eqnarray}
m\dot{v}+\alpha\dot{r} & = & f+\beta \; , \label{eq:Langevin}
\end{eqnarray}
where $m$ is the mass of an object with spatial coordinate $r$, subjected to a force $f$. Linear friction is represented by the constant $\alpha\ge0$, which is related to the thermal fluctuations $\beta$, which can be chosen to be represented by the Gaussian distribution \cite{Parisi}
\begin{eqnarray}
\langle\beta(t)\rangle & = & 0 \\
\langle\beta(t)\beta(t^\prime)\rangle & = & 2\alpha k_BT \delta(t-t^\prime) \; , 
\end{eqnarray}
where $k_B$ is Boltzmann's constant and $T$ is the thermodynamic temperature.

When simulating this equation, we adopt the GJF \cite{GJF1} discrete-time approximation to the Langevin equation, written:
\begin{eqnarray}
r^{n+1} & = & r^n + b [dt\, v^n+\frac{dt^2}{2m}f^n+\frac{dt}{2m}\beta^{n+1}] \label{eq:gjf_r} \\
v^{n+1} & = & a\, v^n+\frac{dt}{2m}(af^n+f^{n+1})+\frac{b}{m}\beta^{n+1} \, , \label{eq:gjf_v}
\end{eqnarray}
where the discrete-time notation $r^n=r(t_n)$ indicates that we only have approximations at times $t_n=t_0+n\,dt$, separated by the time step $dt$. The on-site velocity variable $v^n$ approximates $v(t_n)=\dot{r}(t_n)$, and the discrete-time force is $f^n=f(t_n,r^n)$. The discrete-time fluctuation-dissipation relationship is ensured by the coefficients
\begin{eqnarray}
a & = & \frac{\displaystyle{1-\frac{\alpha dt}{2m}}}{\displaystyle{1+\frac{\alpha dt}{2m}}} \label{eq:a} \\
b & = & \frac{\displaystyle{1}}{\displaystyle{1+\frac{\alpha dt}{2m}}} \, , \label{eq:b}
\end{eqnarray}
with the associated integrated fluctuations
\begin{eqnarray}
\beta^{n+1} & = & \int_{t_n}^{t_{n+1}}\beta(t^\prime)\,dt^\prime \, , \label{eq:discrete_beta}
\end{eqnarray}
which are uncorrelated Gaussian random numbers with zero mean and a variance given by the temperature and friction coefficient:
\begin{eqnarray}
\langle\beta^n\rangle & = & 0 \label{eq:noise_dis_ave} \\
\langle\beta^n\beta^l\rangle & = & 2\alpha k_BT dt \delta_{n,l} \, . \label{eq:noise_dis_std}
\end{eqnarray}
We use the {\it ran3()} random number generator as descibed in Ref.~\cite{NR}.
The equations (\ref{eq:gjf_r}) and (\ref{eq:gjf_v}) can be conveniently written in the compact single time step form
\begin{eqnarray}
u^{n+\frac{1}{2}} & = & \sqrt{b}\;\left[v^n+\frac{dt}{2m}f^n+\frac{1}{2m}\beta^{n+1}\right] \label{eq:2GJ_u}\\
r^{n+1} & = & r^n+\sqrt{b}\,dt\,u^{n+\frac{1}{2}}\label{eq:2GJ_r}\\
v^{n+1} & = & \frac{a}{\sqrt{b}}u^{n+\frac{1}{2}}+\frac{dt}{2m}f^{n+1}+\frac{1}{2m}\beta^{n+1}\label{eq:2GJ_v} \, ,
\end{eqnarray}
where each time step is initiated by $(r^n,v^n)$ and concludes with $(r^{n+1},v^{n+1})$.
The half-step velocity variable $u^{n+\frac{1}{2}}$ \cite{2GJ} approximates $v(t_{n+\frac{1}{2}})=\dot{r}(t_{n+\frac{1}{2}})$, and is found to be thermodynamically robust in its ability to produce reliable kinetic measures. 
Specifically, for linear systems, when $f=-\kappa r$ ($\kappa\ge0$), it has been shown that the trajectory $r^n$ samples the correct configurational Boltzmann statistics \cite{GJF1} and that the half-step velocity defined by Eq.~(\ref{eq:2GJ_r}) is resulting in exact kinetic measures \cite{2GJ} regardless of the applied time step within the stability range. Very robust statistical behavior has been numerically demonstrated for nonlinear and complex systems as well. However, it is known that the on-site velocity variable $v^n$ does not precisely sample the desired Maxwell-Boltzmann statistics for $\kappa>0$ \cite{GJF1}, and a systematic quadratic error in the reduced time step is found as the time step is increased.

We will here be concerned with noticeable statistical discrepancies that have been observed in Langevin simulations of nonlinear systems. In particular, the significant deviations from perfect statistics observed for, e.g., single particle behavior in nonlinear, including corrugated, potentials \cite{GJF4} in the low damping limit, are characterized by very sudden departures from correct statistics at or near relatively large time steps within the stability range. This indicates that the source of the deviation is not one of the usual algorithmic errors, which tend to increase gradually with the time step. While the general feature of the described sudden discrepancies are ubiquitous for the standard Langevin algorithms that we are aware of, including the classic methods described and analyzed in Refs.~\cite{SS,BBK,Pastor_88}, the deviations are particularly noticeable when using the GJF-2GJ algorithm, given its reliable statistical properties for elevated time steps. We here simplify and limit the investigation to a very simple system, the pendulum equation, which we can analyze analytically and numerically in order to discover the reason for the anomalous errors in the otherwise robust simulation environment.

\section{The Pendulum and its discrete-time statistics}
We consider the corrugated periodic potential
\begin{eqnarray}
E_p & = & E_0(1-\cos\frac{r}{r_0})\, , \label{eq:pendulum_pot}
\end{eqnarray}
where $E_0=\kappa r_0^2\ge0$ is a characteristic energy, $r_0$ is a characteristic distance, and $\kappa$ is a constant $\kappa>0$. The resulting force to be inserted into the Langevin equations and the GJF algorithm above is then
\begin{eqnarray}
f & = & -\kappa r_0\sin\frac{r}{r_0}\, . \label{eq:pendulum_force}
\end{eqnarray}
From this potential we define the characteristic time as the inverse of the small-amplitude natural oscillation frequency $\Omega_0=\sqrt{\kappa/m}$ around the stable fixed point at $r=0$ for $\alpha=0$. The characteristic velocity is thus $v_0=r_0\Omega_0$. Given that the largest curvature of the potential is found for $r=0$, we define the relevant overall stability limit for the Verlet-type \cite{Stormer_1921,Verlet} integration method as given by $\Omega_0dt<2$, where $\Omega_0dt$ is the reduced time step. This limitation ensures that the dynamics is stable everywhere on the potential surface, including the ground state, and it ensures that the discrete-time dynamics in the nonlinear regime does not challenge the stability. Thus, algorithmic stability properties are given by the small amplitude linear oscillations.

\begin{figure}[t]
\centering
\scalebox{0.45}{\centering \includegraphics[trim={2cm 3.5cm 1.5cm 3.0cm},clip]{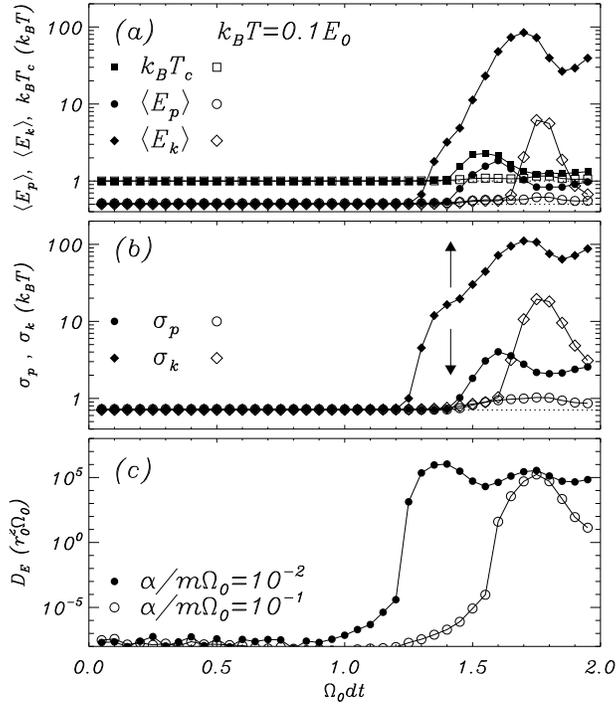}}
\caption{Statistical averages Eq.~(\ref{eq:measure_Ep})-(\ref{eq:measure_DE}) as a function of reduced time step $\Omega_0dt$ for the nonlinear oscillator described by the potential in Eq.~(\ref{eq:pendulum_pot}), and simulated by Eqs.~(\ref{eq:2GJ_u})-(\ref{eq:2GJ_v}). Each marker is the result of 10$^{9}$ simulated time steps for 1,000 pendula, and the parameter values represented by the markers are indicated on the figures. Open markers represent $\alpha/m\Omega_0=10^{-2}$, filled markers represent $\alpha/m\Omega_0=10^{-1}$. Simulated thermal energy is $k_BT=0.1E_0$. Vertical arrows point to $\Omega_0dt=\sqrt{2}$. Horizontal dotted lines are given at the expected thermodynamic values $T_c=2\langle E_k\rangle=k_BT$ (a), and $\sigma_k=k_BT/\sqrt{2}$ (b).
}
\label{fig_1}
\end{figure}

\begin{figure}[t]
\centering
\scalebox{0.45}{\centering \includegraphics[trim={2cm 3.5cm 1.5cm 3.0cm},clip]{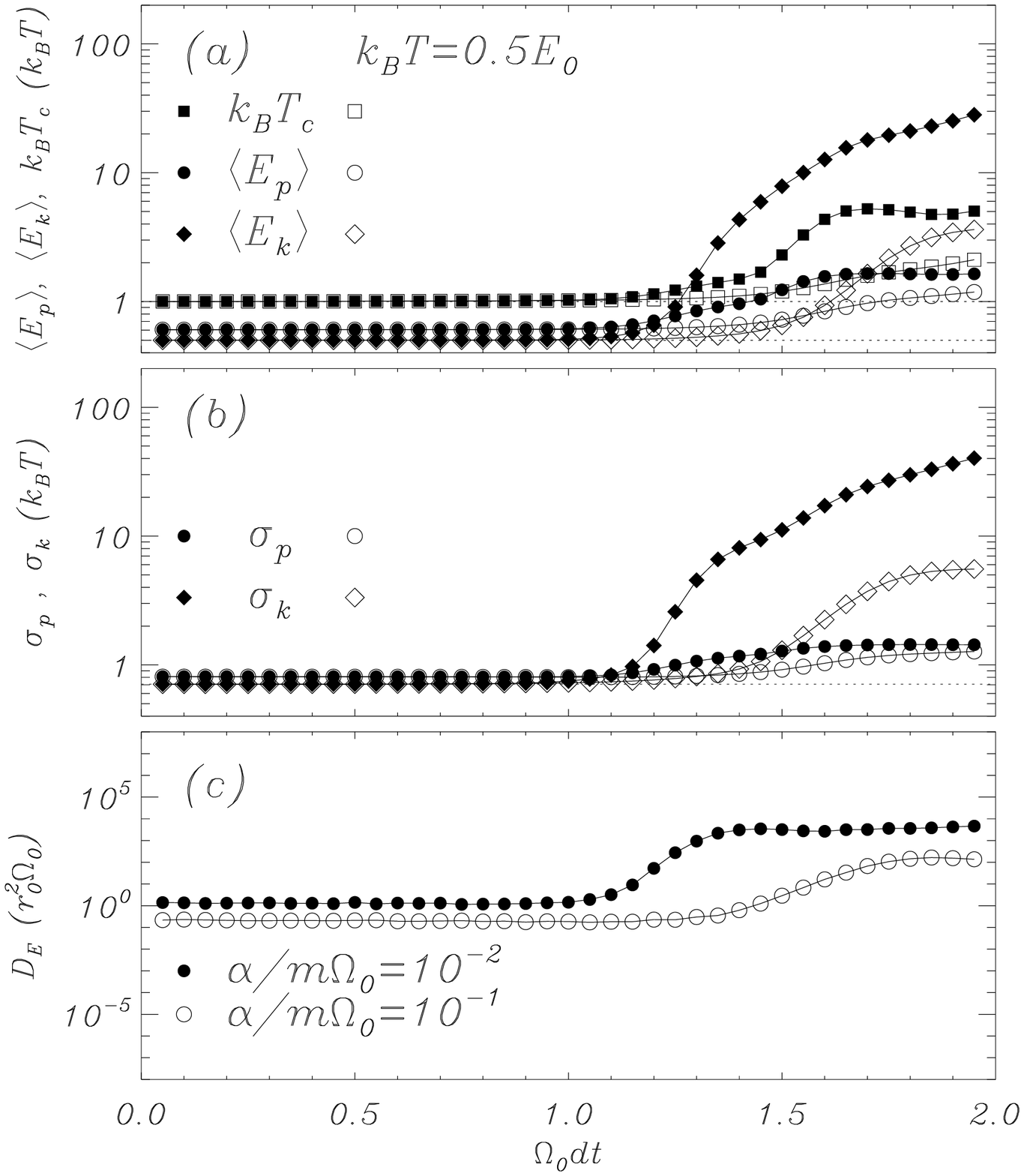}}
\caption{Statistical averages Eq.~(\ref{eq:measure_Ep})-(\ref{eq:measure_DE}) as a function of reduced time step $\Omega_0dt$ for the nonlinear oscillator described by the potential in Eq.~(\ref{eq:pendulum_pot}), and simulated by Eqs.~(\ref{eq:2GJ_u})-(\ref{eq:2GJ_v}). Each marker is the result of 10$^{9}$ simulated time steps for 1,000 pendula, and the parameter values represented by the markers are indicated on the figures. Open markers represent $\alpha/m\Omega_0=10^{-2}$, filled markers represent $\alpha/m\Omega_0=10^{-1}$. Simulated thermal energy is $k_BT=0.5E_0$.  Horizontal dotted lines are given at the expected thermodynamic values $T_c=2\langle E_k\rangle=k_BT$ (a), and $\sigma_k=k_BT/\sqrt{2}$ (b).
}
\label{fig_2}
\end{figure}

The phenomenon under investigation is visible from the simulation results shown in Figures~\ref{fig_1} and \ref{fig_2}. These data points are derived from simulating 1,000 independent pendula for each set of parameter values \cite{comment_on_1000}. Each data point is generated from first equilibrating each pendulum for a normalized time of at least 10$\times$$m\Omega_0/{\alpha}$ before making statistical averages for the next 10$^8$ time steps. After this, the time step is increased slightly to make the next set of data points. This is done using the algorithm Eqs.~(\ref{eq:2GJ_u})-(\ref{eq:2GJ_v}). We produce the statistics for the following quantities of energy, energy fluctuations, and diffusion.
The displayed measures are
\begin{eqnarray}
\langle E_p \rangle & = & \langle E_p(r^n) \rangle \label{eq:measure_Ep}\\
\langle E_k \rangle & = & \frac{1}{2}k_BT_k \; = \; \frac{1}{2}m\langle u^{n+\frac{1}{2}}u^{n+\frac{1}{2}}\rangle\label{eq:measure_Ek}\\
k_BT_c & = & \frac{\langle (E_p^\prime(r^n))^2\rangle}{\langle E_p^{\prime\prime}(r^n)\rangle}\label{eq:measure_Tc}\\
\sigma_p^2 & = & \langle (E_p(r^n))^2\rangle-\langle E_p(r^n)\rangle^2\label{eq:measure_sp}\\
\sigma_k^2 & = & \langle (E_k(u^{n+\frac{1}{2}}))^2\rangle-\langle E_k(u^{n+\frac{1}{2}})\rangle^2\label{eq:measure_sk}\\
D_E & = & \lim_{n\rightarrow\infty}\frac{\langle(r^{q+n}-r^q)^2\rangle_q}{2\,dt \, n}\label{eq:measure_DE} \, ,
\end{eqnarray}
where $\langle E_p \rangle$ and $\langle E_k \rangle$ are the mean values of potential and kinetic energy, respectively; $\sigma_p$ and $\sigma_k$ are the associated energy fluctuations; and $k_BT_c$ is the energy of the configurational temperature \cite{Tc}. If the sampled velocities are given by a correct Maxwell-Boltzmann distribution then the kinetic fluctuation is expected to be $\sigma_k=k_BT/\sqrt{2}$. The diffusion $D_E$ is measured by the Einstein definition \cite{AllenTildesley} (see comments on the discrete-time relationship between this definition and the Green-Kubo approximation in Ref.~\cite{2GJ}). Results are shown for representative values of both damping and temperature as a function of the reduced time step $0<\Omega_0dt<2$.

The summarized results shown in Figures~\ref{fig_1} and \ref{fig_2} indicate that the statistical response of the algorithm is predictively robust for all small to moderately sized time steps with time-step-independent results, regardless of the temperature and the damping coefficient. However, sudden onsets of discrepancy for the smaller of the time steps are observed for all statistical measures with the lower damping showing signs of discrepancy at smaller time steps than the simulations for higher damping. The vertical logarithmic scale demonstrates how dramatic these discrepancies are, especially in light of the expected time-step independent behavior that is characteristic for the GJF-2GJ method in both configurational and kinetic sampling.
The following section will propose two types of intrinsic dynamic modes that are responsible for this behavior.

\section{Energetic Modes in Damped Oscillators}
In order to investigate the existence and stability of specific modes, we use the SV form of the GJF method \cite{GJF2} without noise
\begin{eqnarray}
r^{n+1} & = 2br^n-ar^{n-1}+\frac{b\,dt^2}{m}f^n\, , \label{eq:gjf_sv}
\end{eqnarray}
which for the pendulum equation reads
\begin{eqnarray}
r^{n+1} & = 2br^n-ar^{n-1}-b\,\Omega_0^2dt^2r_0\sin\frac{r^n}{r_0} \, . \label{eq:gjf_sv_pend}
\end{eqnarray}
This equation is the discrete-time representation of a purely damped system for $\alpha dt>0$, and the linear limit for the potential $E_p$ is given near the static ground state fixed point $\varepsilon^n\approx r^*=0$ ($|\varepsilon^n|\ll r_0$)
\begin{eqnarray}
\varepsilon^{n+1} & = & 2b(1-\frac{\Omega_0^2dt^2}{2})\varepsilon^n-a\varepsilon^{n-1} \, , \label{eq:discr_damp}
\end{eqnarray}
for which the solutions to the characteristic polynomial confirms the stability of the fixed point $r_0^*=0$ in the entire range $0\le\Omega_0dt<2$. The analyses that led to both the time-step-independent configurational \cite{GJF1} and kinetic \cite{2GJ} results have assumed fluctuations around this fixed point, since this is the physical minimum of the potential. However, given the observations of onsets of significant deviations, as illustrated in Figs.~\ref{fig_1} and \ref{fig_2}, it is reasonable to explore if other modes exist.

Focusing on the results of the low damping value in Figure \ref{fig_1}, we notice that the sudden rise in statistical values is accompanied by a dramatic increase in the diffusion coefficient. This increase happens for relatively large time steps, so it is not surprising that spurious events in a noisy environment can make the system move from one potential well to the next in one time step. However, since the onset of discrepancies is sudden, and the diffusion coefficient becomes {\it very} large, we suspect that more intrinsic behavior is at play.\\

\begin{figure}[t]
\centering
\scalebox{0.45}{\centering \includegraphics[trim={2cm 2.0cm 1.5cm 3.0cm},clip]{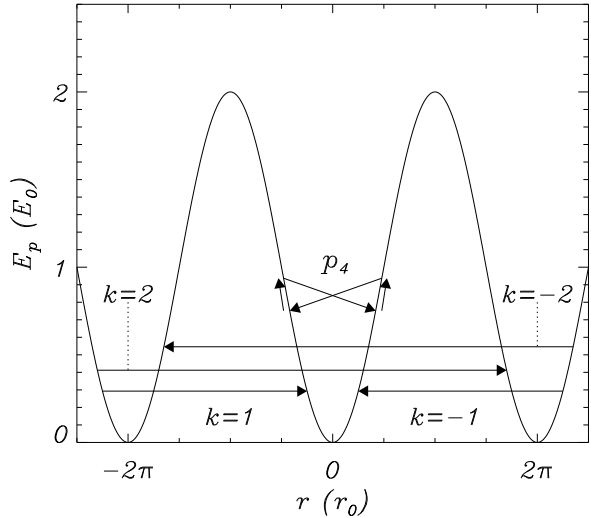}}
\caption{Sketch of the periodic potential and the dynamic ghost modes under investigation. Each arrow indicates the motion of a single time step $dt$ in a mode. The traveling mode is indicated for $k=\pm1,\pm2$ by horizontal arrows, and $p_4$ is the oscillating mode.
}
\label{fig_3}
\end{figure}

\noindent
{\underline{Traveling Mode.}} The first mode we investigate is the traveling solution (sketched in Fig.~\ref{fig_3}) $r^n=r^*+2\pi r_0 k\, n$, where $r^*$ is a constant and $k$ a non-zero integer. This mode travels $k$ potential wells every time step, and it makes contact with the potential at one value $r^*$ in each well.
Inserting this traveling solution into Eq.~(\ref{eq:gjf_sv_pend}) gives
\begin{eqnarray}
2\pi k \frac{\alpha dt}{m} & = & -\Omega_0^2dt^2\sin\frac{r^*}{r_0}\, .
\end{eqnarray}
Thus, a traveling solution with
\begin{eqnarray}
\sin\frac{r^*}{r_0} & = & -\frac{\alpha}{m\Omega_0}\frac{2\pi k}{\Omega_0dt} \label{eq:sol_trav}
\end{eqnarray}
if
\begin{eqnarray}
\Omega_0dt & \ge & 2\pi k\frac{\alpha}{m\Omega_0}\, . \label{eq:cond_trav}
\end{eqnarray}
The existence of this mode is supported by low damping and large time steps such that the dissipation induced from one time step to the next is compensated by making contact with the potential surface only at points $r^*$ where the force perpetuates the motion. This is a mode that exists only in discrete time, as can be seen from Eq.~(\ref{eq:cond_trav}), and it is, of course, unphysical. The stability of the mode is investigated by inserting $r^n=r^*+2\pi r_0 k\, n + \varepsilon^n$ for $|\varepsilon^n|\ll r_0$ into Eq.~(\ref{eq:gjf_sv_pend}), with $r^*$ given by Eq.~(\ref{eq:sol_trav}). The resulting equation for the small perturbation $\varepsilon^n$ is then
\begin{eqnarray}
\varepsilon^{n+1} & = & 2b(1-\frac{\Omega_0^2dt^2}{2}\cos\frac{r^*}{r_0})\varepsilon^n-a\varepsilon^{n-1}\, ,
\end{eqnarray}
which yields the solutions $\lambda_\pm$ to the characteristic polynomial
\begin{eqnarray}
\lambda_\pm & = & b(1-\frac{\Omega_0^2dt^2}{2}\cos\frac{r^*}{r_0}) \pm\sqrt{b^2(1-\frac{\Omega_0^2dt^2}{2}\cos\frac{r^*}{r_0})^2-a} \, .
\end{eqnarray}
It is straightforward to see that for
\begin{eqnarray}
0 & < & b^2\left(1-\frac{\Omega_0^2dt^2}{2}\cos\frac{r^*}{r_0}\right)^2 \; < \; a
\end{eqnarray}
$\lambda_\pm$ are complex and $|\lambda_\pm|^2=a<1$. 
Equally straightforward, albeit more cumbersome, calculations show that $|\lambda_\pm|<1$ for all values $\cos\frac{r^*}{r_0}>0$. The traveling mode is therefore stable for all fixed points in the interval $0\le r^*<\frac{\pi}{2}r_0$. Notice that Eq.~(\ref{eq:cond_trav}) states that the traveling mode can only exist for damping values $\pi k \, \alpha/m\Omega_0<1$. Thus, any stable and self-sustaining traveling mode can exist only for $\alpha/m\Omega_0<1/k\pi$. We emphasize that this kind of mode is truly unusual and unphysical: Not only can the mode be traveling at certain high velocities in a purely damped system, it may also perpetually travel uphill if a potential tilt of limited slope is added to the corrugated potential.

In order to see if this spurious discrete-time mode is related to the curious large-time-step discrepancies observed in Figs.~\ref{fig_1} and \ref{fig_2}, we select a few characteristic values of the reduced time step and investigate the velocity density distribution $\rho(u)$ acquired from the simulations in Figs.~\ref{fig_1} and \ref{fig_2}. The statistical expectation of this distribution $\rho_k(u)$ is the Maxwell-Boltzmann distribution
\begin{eqnarray}
\rho_k(u) & \propto & \exp(-\frac{\frac{1}{2}mu^2}{k_BT}) \, .
\end{eqnarray}
Thus, we define an effective kinetic potential $U_{MB}(u)$ from the simulated density distribution $\rho_k(u)$
\begin{eqnarray}
U_{MB}(u) & = & -k_BT\ln\rho_k(u) + {\cal C}_k \, , \label{eq:MB}
\end{eqnarray}
with ${\cal C}_k=k_BT\ln\rho_k(0)$
such that $U_{MB}(u)=\frac{1}{2}mu^2$, if the statistics is in accordance with thermal physics.

\begin{figure}[t]
\centering
\scalebox{0.45}{\centering \includegraphics[trim={1.5cm 2.0cm 2.0cm 3.0cm},clip]{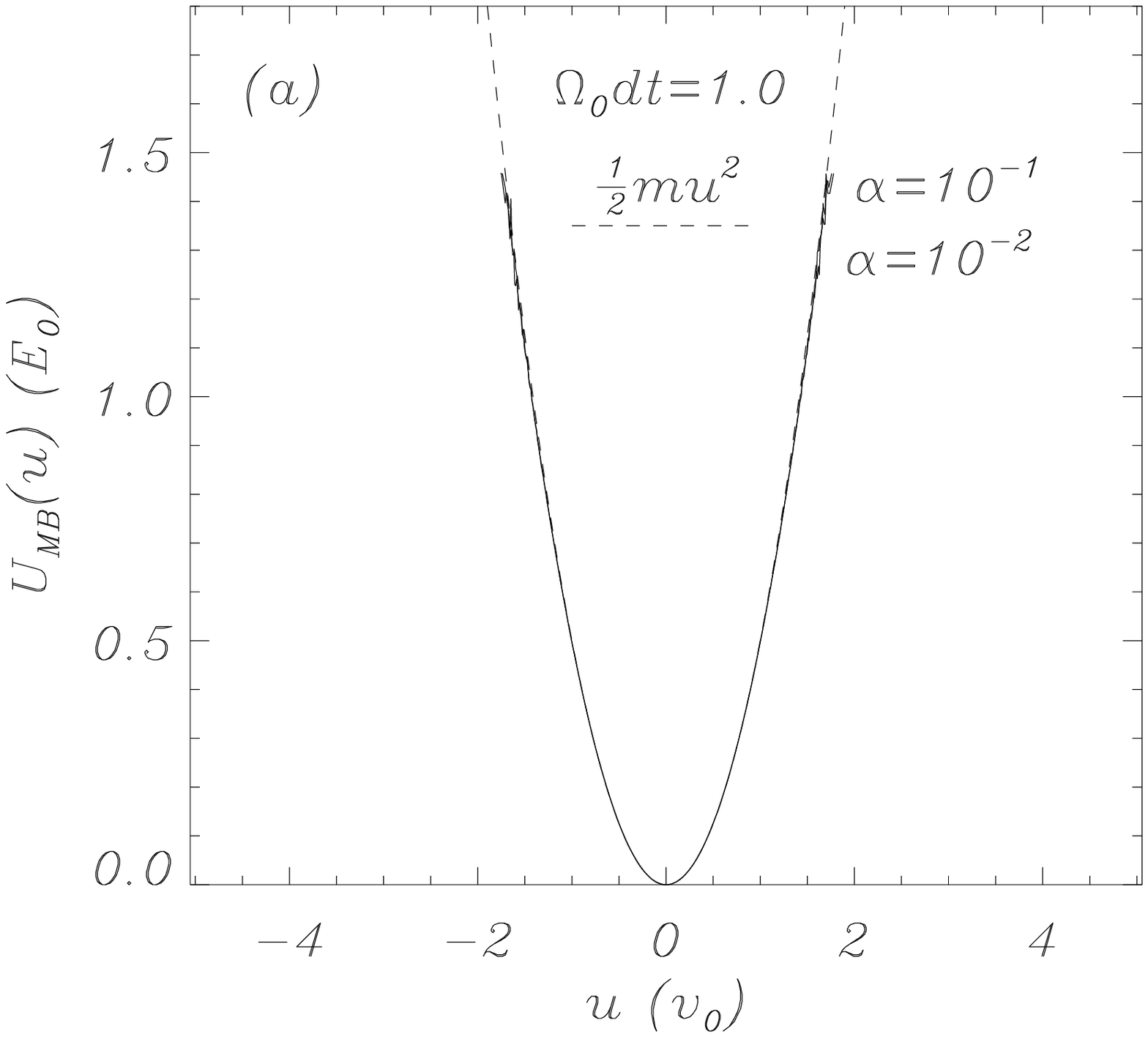}}
\scalebox{0.45}{\centering \includegraphics[trim={1.5cm 2.0cm 2.0cm 3.0cm},clip]{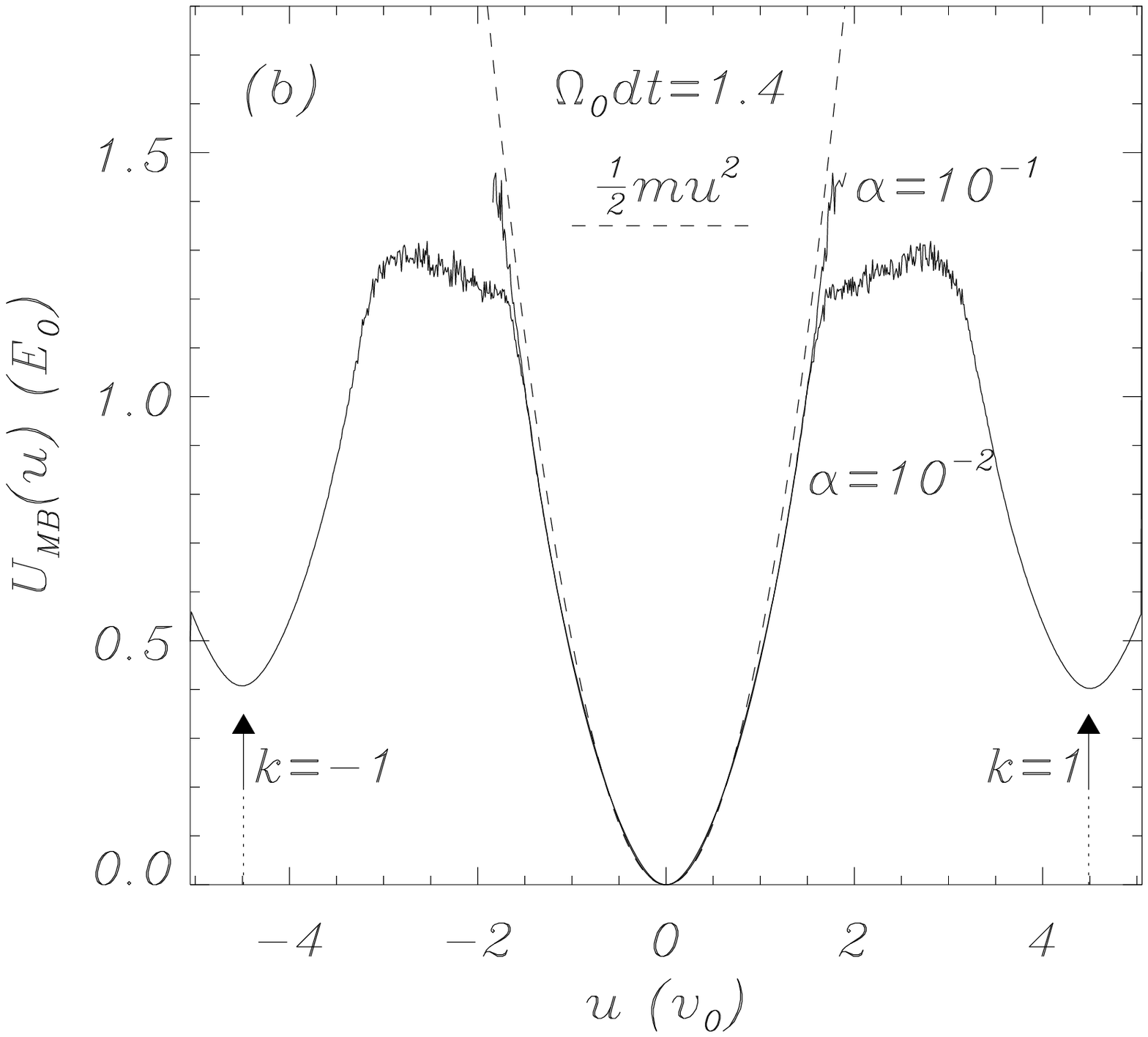}}
\scalebox{0.45}{\centering \includegraphics[trim={1.5cm 2.0cm 2.0cm 3.0cm},clip]{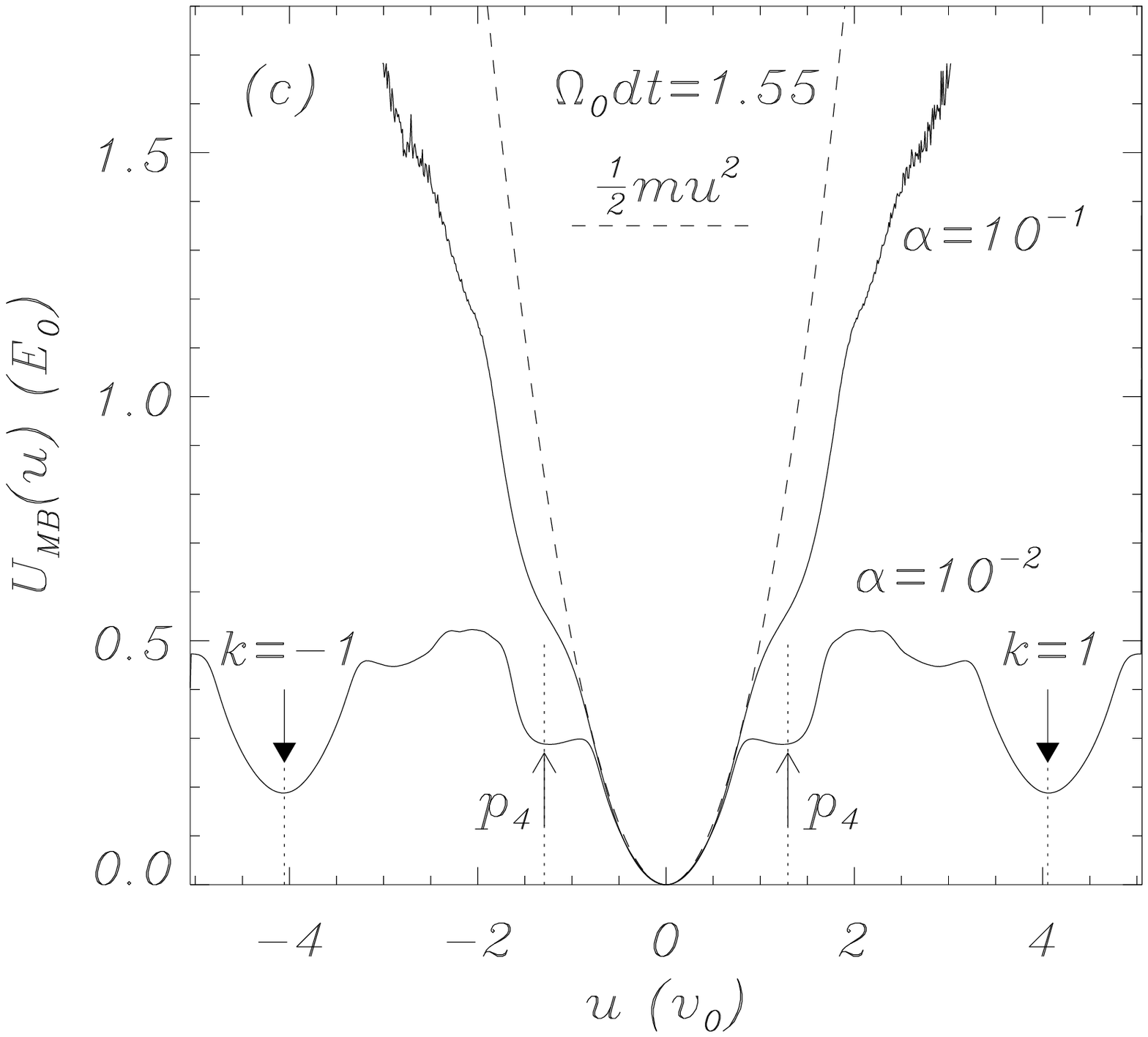}}
\scalebox{0.45}{\centering \includegraphics[trim={1.5cm 2.0cm 2.0cm 3.0cm},clip]{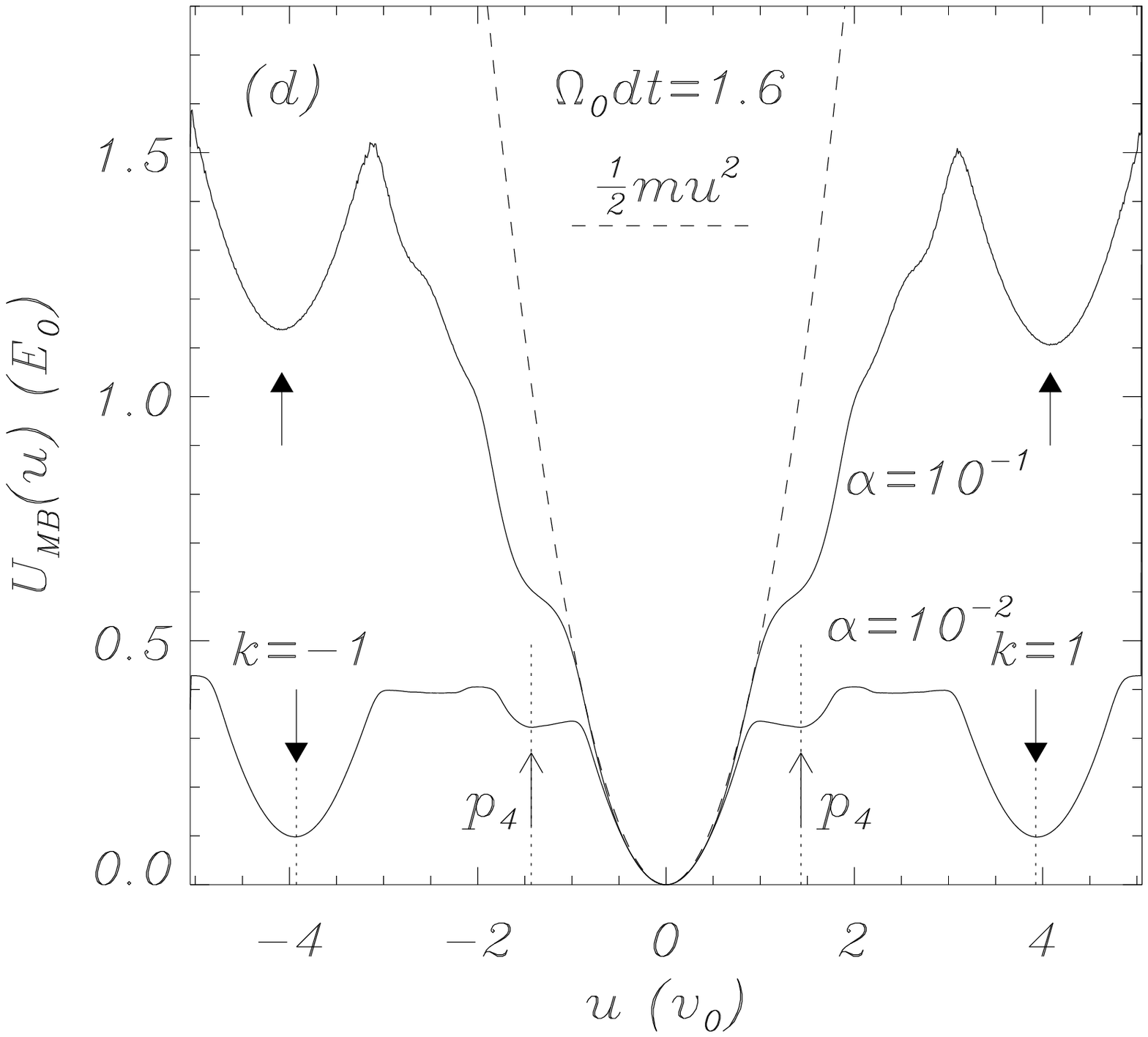}}
\caption{Effective kinetic potential $U_{MB}(u)$ derived from acquired velocity distribution functions (solid curves). Results for two values of damping coefficient $\alpha/m\Omega_0$ are shown. Results for larger damping are above those of lower damping unless the curves coincide. The ideal kinetic potential is indicated by the dashed curve. $k_BT=0.1E_0$. Other parameters given on the figure.  Vertical arrows indicate characteristic velocities of intrinsic dynamical modes discussed in the text. $k=\pm1$ refers to the traveling mode  with velocities given by Eq.~(\ref{eq:v_k}), and $p_4$ in (c) and (d) refers to the oscillating mode with velocities given by Eq.~(\ref{eq:vel_p4}).
}
\label{fig_4}
\end{figure}

Figure \ref{fig_4} shows the comparisons for $\alpha/m\Omega_0=0.01$ and $\alpha/m\Omega_0=0.1$ for different time steps. In accordance with Fig.~\ref{fig_1}, we expect that $\Omega_0dt=1$ will show a physically meaningful distribution since all the statistical values in Fig.~\ref{fig_1} are time-step independent up to at least $\Omega_0dt\approx1$. That is confirmed by Fig.~\ref{fig_4}a, which shows perfect agreement between $U_{MB}(u)$ and $\frac{1}{2}mu^2$ for both values of damping. Thus, the algorithm samples a physically meaningful kinetic distribution even for this high time step, as expected from Ref.~\cite{2GJ}.

For $\Omega_0dt=1.4$, Fig.~\ref{fig_1} shows that the low damping case $\alpha/m\Omega_0=0.01$ significantly deviates from the physical result, while the simulation for the damping value $\alpha/m\Omega_0=0.1$ is still in accordance with the statistical expectation. Figure~\ref{fig_4}b shows the details of this result. While the distributions are in perfect agreement with physical expectations for all velocities up to a certain value $u\approx1.5v_0$, significant large-velocity departure from the quadratic behavior is noticeable in $U_{MB}(u)$ for $\alpha/m\Omega_0=0.01$. Further, the departure is characterized by a particular speed, which coincides precisely with the velocity of the noiseless traveling mode
\begin{eqnarray}
v_k & = & \frac{2\pi k}{\Omega_0dt} v_0 \, . \label{eq:v_k}
\end{eqnarray}
This velocity is indicated on Figs.~\ref{fig_4}bcd by arrows for $k=\pm1$. Thus, we conclude that the departure from physical statistics is due to this intrinsic dynamical mode and not a systematic error that increases by some power of the reduced time step $\Omega_0dt$. This indicates that the algorithm is sampling the phase space correctly, but that the phase space for $\Omega_0dt$ larger than a certain threshold contains an expanded set of possible states, which the algorithm is statistically sampling. However, Fig.~\ref{fig_4} shows that for as long as the system is sampling the behavior around $u=0$ then the statistics is in agreement with the Maxwell-Boltzmann distribution. The observation that the damping value $\alpha/m\Omega_0=0.1$ does not show departure from physical statistics in Fig.~\ref{fig_1} for $\Omega_0dt=1.4$ is corroborated by the perfect distribution function seen in Fig.~\ref{fig_4}b.\\

\noindent
{\underline{Oscillating Mode.}} For higher values of $\Omega_0dt$ we see from Fig.~\ref{fig_1} that both damping values $\alpha/m\Omega_0=0.01$ and $\alpha/m\Omega_0=0.1$ show enhanced statistical averages. Figure~\ref{fig_4}c shows that while the distribution for the higher damping indicates that the traveling mode has not been sampled, the distribution abruptly separates from the physical one (dashed) for $|u|>1$, when also the distribution for low damping separates. Given that this is a distinct feature commensurate with neither the physical nor the traveling mode, we explore other options.

It is tempting to suggest a period two ($p_2$) oscillating mode of the kind $r^{n+1}=-r^{n}=r^*>0$ as an inherent dynamical mode in a convex potential. Inserting this form into Eq.~(\ref{eq:gjf_sv_pend}) shows the following:
\begin{eqnarray}
r^* & = & -2br^*-ar^*+b\Omega_0^2dt^2r_0\sin\frac{r^*}{r_0} \\
\Rightarrow \; \; \frac{\sin\frac{r^*}{r_0}}{\frac{r^*}{r_0}} & = & \frac{4}{\Omega_0^2dt^2} \, .
\end{eqnarray}
Since this is not possible within the defined stability limit $\Omega_0dt<2$, we conclude that this mode does not exist.

Instead we propose the $p_4$ oscillatory mode sketched in Fig.~\ref{fig_3}, where $r^{n+1}=r^*_1=-r^{n-1}>0$ and $r^n=r^*_2=-r^{n-2}>0$ with $r^*_1\ge r^*_2$. Inserting this mode into Eq.~(\ref{eq:gjf_sv_pend}) yields the two equations
\begin{eqnarray}
 \frac{\alpha dt}{2m}r^*_1 & = & r^*_2-\frac{\Omega_0^2dt^2}{2}r_0\sin\frac{r^*_2}{r_0} \label{eq:p4_1}\\
-\frac{\alpha dt}{2m}r^*_2 & = & r^*_1-\frac{\Omega_0^2dt^2}{2}r_0\sin\frac{r^*_1}{r_0} \label{eq:p4_2}\, .
\end{eqnarray}
For $\alpha=0$ we can see that the two equations decouple, and the common solution $r_1^*=r_2^*=r^*$ is given by
\begin{eqnarray}
\frac{\sin\frac{r^*}{r_0}}{\frac{r^*}{r_0}} & = & \frac{2}{\Omega_0^2dt^2}\, .
\end{eqnarray}
Thus, it is possible to find such fixed points for $\Omega_0dt>\sqrt{2}$. Approximating the left hand side of this equation with its quadratic polynomial expansion around $r^*=0$, we can make a convenient expression of the solution
\begin{eqnarray}
r^* & \approx & \sqrt{6(1-\frac{2}{\Omega_0^2dt^2})}\, r_0 \, . \label{eq:r_i_approx}
\end{eqnarray}
For $0<\alpha/m\Omega_0$, we see from Eqs.~(\ref{eq:p4_1}) and (\ref{eq:p4_2}) that the two values $r_1^*$ and $r_2^*$ split such that, to first order in $\alpha dt/2m$,
\begin{eqnarray}
r^*_i &\approx & r^*(1\pm\frac{\frac{\alpha dt}{2m}}{\frac{\Omega_0^2dt^2}{2}\cos\frac{r^*}{r_0}-1}) \, ,
\end{eqnarray}
which is an expression that must be applied for $\Omega_0dt>\sqrt{2}$ and we must further require that $\cos\frac{r^*}{r_0}>2/\Omega_0^2dt^2$. Applying the approximation Eq.~(\ref{eq:r_i_approx}) gives
\begin{eqnarray}
r_i^* & \approx & r^* \, (1\pm\frac{\frac{\alpha dt}{2m}}{5-\Omega_0^2dt^2})\\
& = &  \sqrt{6(1-\frac{2}{\Omega_0^2dt^2})} \, r_0 \, (1\pm\frac{\frac{\alpha dt}{2m}}{5-\Omega_0^2dt^2}) \, .
\end{eqnarray}

The resulting $p_4$ velocities stemming from this oscillating $p_4$ mode are then
\begin{eqnarray}
v^{n+\frac{1}{2}} & = & \frac{r^{n+1}-r^n}{dt} \; = \;\pm\left\{\begin{array}{c} \displaystyle\frac{r^*_1-r^*_2}{dt} \\ \displaystyle\frac{r^*_1+r^*_2}{dt}\end{array}\right. \, , \label{eq:vel_p4}
\end{eqnarray}
where $v=0$ and $v=2r_i^*/dt$ for $\alpha=0$. We have indicated the large magnitude $p_4$ mode velocities in Figure~\ref{fig_4}c, and we see that the observation that led to the search for a non-traveling mode is entirely consistent with this oscillatory mode for both values of simulated damping (notice that we have not indicated the velocity at $v\approx0$, since that statistical contribution is insignificant compared to the vast majority of events being found near $v=0$).

Figure~\ref{fig_4}d shows the effective kinetic potential derived from the velocity density distributions for a time step value, where we observe traveling mode contributions for both simulated damping values \cite{sqrt_b}. Again, we see complete consistency between the anomalies in the kinetic distributions and the identified modes. We also reemphasize that the overall sampling of the low velocity regime (the regime characterized by velocities lower than the activated dynamic modes) is largely unaffected by the existence of the large velocity (nonlinear) modes. This indicates that the algorithm is performing very reasonable statistical sampling of the modes that are present in the discrete-time system. It is important to note that the two identified modes are not suggested to always be the only relevant ones. Instead, we expect that other intrinsically dynamical modes may play a role in corrupting the expected thermal statistics stemming from such large time step simulations at low damping and non-vanishing temperature. In fact, the shown distributions (effective kinetic potentials) show fine structures that may suggest exactly that.

\begin{figure}[t]
\centering
\scalebox{0.45}{\includegraphics[trim={1.5cm 2.0cm 2.0cm 3.0cm},clip]{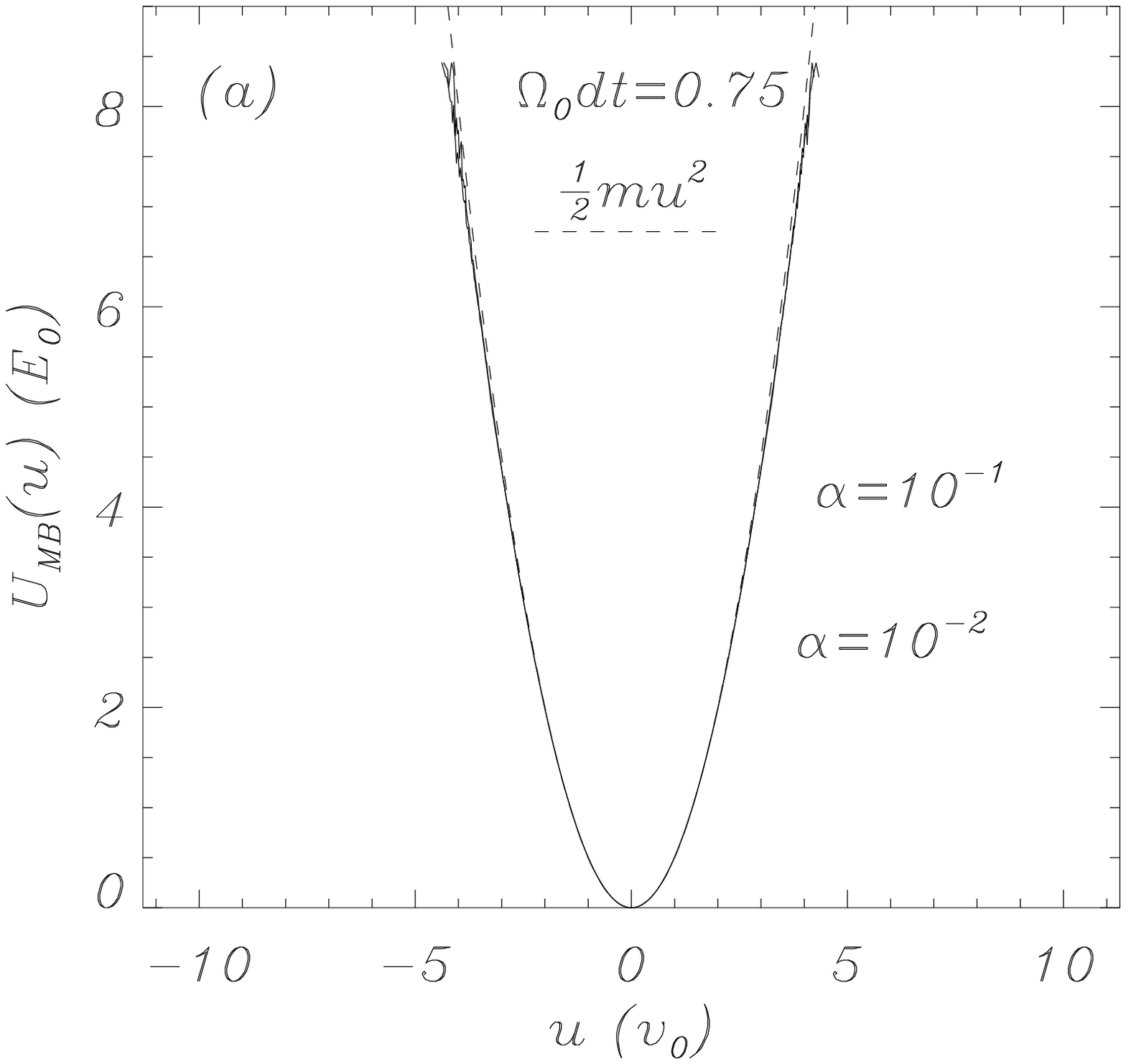}}
\scalebox{0.45}{\includegraphics[trim={1.5cm 2.0cm 2.0cm 3.0cm},clip]{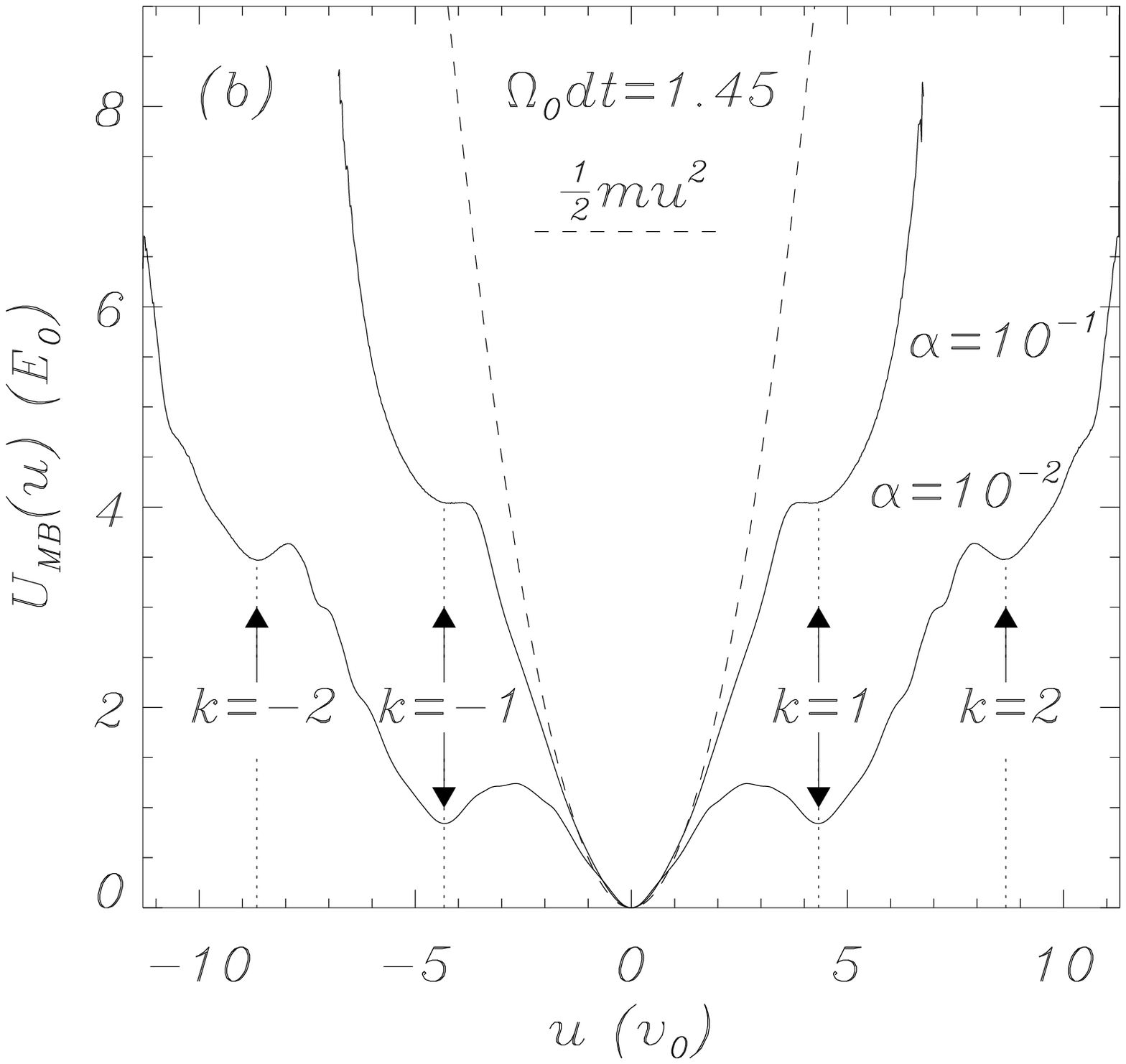}}
\caption{Effective kinetic potential $U_{MB}(u)$ derived from acquired velocity distribution functions (solid curves). Results for two values of damping coefficient $\alpha/m\Omega_0$ are shown. Results for larger damping are above those of lower damping, unless the curves coincide. The ideal kinetic potential is indicated by the dashed curve. $k_BT=0.5E_0$. Other parameters given on the figure. Vertical arrows indicate characteristic velocities of intrinsic dynamical modes discussed in the text. $k=\pm1$ and $k=\pm2$ refer to the traveling modes with velocities given by Eq.~(\ref{eq:v_k}).}
\label{fig_5}
\end{figure}

The consistency between the data and the identified modes is further exemplified by viewing results for the much higher thermal energy $k_BT=0.5E_0$. While Figure~\ref{fig_5}a shows the distributions for $\Omega_0dt=0.75$, where both damping values result in expected thermal statistics based on the data in Figure~\ref{fig_2}, Figure~\ref{fig_5}b (simulated at $\Omega_0dt=1.6$) displays not only the $k=\pm1$ traveling modes as a reason for the enhanced statistical measures, but also visible signatures of the $k=\pm2$ mode for the lower of the two damping values.

\begin{figure}[t]
\centering
\scalebox{0.45}{\includegraphics[trim={1.5cm 2.0cm 2.0cm 3.0cm},clip]{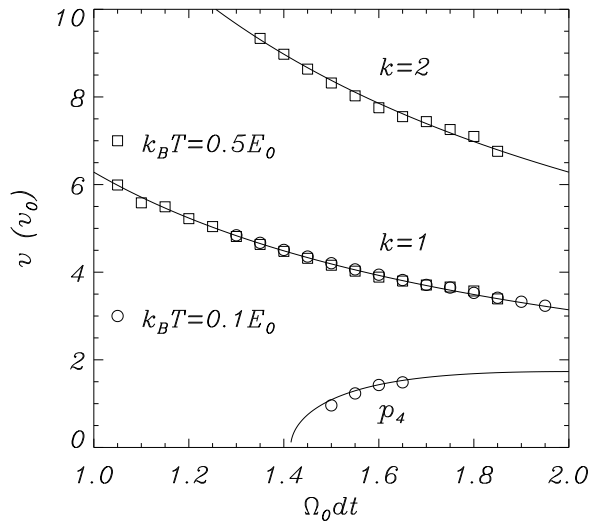}}
\caption{Identifiable local minima in $U_{MB}(u)$ (markers) as a function of applied time step $\Omega_0dt$. $\alpha/m\Omega_0=0.01$. Solid curves represent non-thermal velocities for the traveling modes $k=1,2$ (Eq.~(\ref{eq:v_k})) and for the $p_4$ mode (Eq.~(\ref{eq:vel_p4})).}
\label{fig_6}
\end{figure}

Figure~\ref{fig_6} displays direct comparisons between the predicted velocities of the traveling and $p_4$ modes, as given by Eqs.~(\ref{eq:v_k}) and (\ref{eq:vel_p4}), and results obtained from locating the visible local minima of $U_{MB}(u)$ in figures of the kind shown in Figs.~\ref{fig_4} and \ref{fig_5}. Notice that the data shown in Fig.~\ref{fig_6} are only cases where we can locate an actual minimum; thus, the effect of, e.g., the $p_4$ mode is much broader than what this figure indicates. However, the plot clearly shows that the identified modes are in excellent agreement with the identifiable depressions in the effective kinetic potential $U_{MB}(u)$, and we conclude that both the traveling and $p_4$ modes are major contributors to the departure from the physical Maxwell-Boltzmann statistics observed for this system.

We submit that the discrepancies in the expectations of statistical response observed in Figures~\ref{fig_1} and \ref{fig_2} are not due to systematic algorithmic time-step errors, but instead are related to the inherent properties of discrete time, which possesses nonlinear, unphysical and modes that can be activated by the thermal sampling. The energetics of these ghost modes are therefore relevant components to understand if and when these modes may become important for the results. For example, we have found that the traveling modes exist (and are stable) in the entire time step region given by Eq.~(\ref{eq:cond_trav}), which means that it could be observed for $\Omega_0dt>0.0628$ ($k=1$ for $\alpha/m\Omega_0=0.01$) and $\Omega_0dt>0.628$ ($k=1$ for $\alpha/m\Omega_0=0.1$). Yet, they are activated for much larger time steps. The reason is likely that while these modes are possible, they have an energy $E$ (mostly kinetic), inversely proportional to the square of the time step. Thus, it is extremely implausible to reach these states for small $\Omega_0dt$, and if these states are reached then they may only have limited life time, given the thermal properties of the system. However, for larger time steps, this discrete-time energy barrier between the physical ground state and one of the ghost modes decreases, and it becomes more likely to be temporarily trapped in such mode. An interesting detail seen in Figure~\ref{fig_1} is the coexistence of different ghost modes for $\alpha/m\Omega_0=0.01$ (filled markers). The onset of the traveling mode is clearly seen by the rise in all kinetic measures at $\Omega_0dt\ge1.3$. In contrast, the potential energy and its fluctuations remain depressed up through $\Omega_0dt\le1.4$. This is consistent with the traveling mode for very low damping, since Eq.~(\ref{eq:sol_trav}) indicates that the potential energy contribution from the traveling mode ($k=\pm1$) is $E_p\approx0.002E_0$. Thus, this is insignificant compared to the contribution from the thermal bath. However, the rise in the potential energy seen for $\Omega_0dt>1.4$ coincides with the emergence of the $p_4$ mode, which we have seen in Figure~\ref{fig_4}c (and not in Figure~\ref{fig_4}b). The $p_4$ mode has an average energy (given by Eq.~(\ref{eq:r_i_approx})) of $E_p\approx0.5E_0$ for $\Omega_0dt=1.55$. With the thermal potential energy being less than an order of magnitude below that value, the emergence of the $p_4$ mode is noticed in the statistics of the potential energy, even if the kinetic measures are dominated by the traveling mode contributions.

We have given examples of some of the modes possible in this system. However, as mentioned above, more modes may be relevant, and still others can be found for other types of nonlinear systems. The message of this presentation is to focus on the phenomenon of inherently stable, inherently unphysical, large time step modes interfering with the physical statistics under investigation. We also emphasize the difference between this kind of deviation from correct statistics and the usual systematic errors that are normally expected from numerical algorithms.

\section{Discussion}
\label{sec_lab_IV}
The discrete-time ghost modes discussed in this paper have been shown to be closely aligned with the discrepancies observed between simulated thermodynamics for large time steps and the physical expectations of the continuous-time system of interest, especially those with low characteristic damping. Through detailed investigation of the velocity density distributions obtained from discrete-time simulations, we have been able to directly identify signatures of ghost modes in those distributions, and thereby correlate the sampling of these modes with the observed deviations in other statistical measures, such as averages and fluctuations of energies as well as diffusion. Interestingly, we observe that the velocity density distributions are in near perfect agreement with the Maxwell-Boltzmann distribution for low velocities, even if the high velocity contributions are corrupted by the discrete-time ghost modes. This indicates that the GJF-2GJ method correctly samples the part of the phase space that behaves physically correct, and that the deviations can be attributed to identifiable discrete-time resonances instead of a gradual, systematic increase of algorithmic approximation errors expected from other algorithms. The system chosen for this study has been the simple pendulum, since this is a prototypical system in nonlinear dynamics with both potential wells and possible diffusion. The most dominant mode that influences this system seems to be the traveling mode, which is unique to potentials of limited magnitude. However, the other mode (oscillating $p_4$), which we have explored and identified in the velocity distribution function, is representative of modes relevant for confining potentials in which unlimited diffusion or transport is impossible. In fact, we have confirmed that this kind of mode, as well as other modes with higher periodicity (not discussed above), are sources of statistical errors in confined nonlinear systems, consistent with the Molecular Dynamics observations in Ref.~\cite{Schlick_98,Ma}. Thus, the results put forward in this paper seem to apply much more generally than just the pendulum model.

One can ask if the existence of ghost modes for a given time step puts simulations into danger, and if one should always test for these modes before conducting a simulation. For example, for low damping, we have seen that the traveling ghost modes both exist and are stable even for very low time steps, where one often conducts otherwise reliable simulations. The reason that the ghost modes do not typically interfere with the simulation results for smaller time steps is that they are energetically separated from the physical modes by such a large amount that the probability of being transitioned into a ghost mode is practically infinitesimal. As we have seen, however, it is possible to experience these exclusively discrete-time modes under conditions of, e.g., high temperature or larger time steps. A hint from this paper is that one can use the velocity density distribution function to accurately identify that such high energy, high velocity resonant mode has been caught by the sampling. This opportunity is presented by the GJF-2GJ algorithm since the 2GJ velocity is shown to be statistically correct for as long as the simulation is sampling the physical ground state \cite{2GJ}.

We reemphasize that the resonant discrete-time ghost modes are not unique to the GJF algorithm. In fact, because the origin of the discrepancies are not rooted in linear stability, but instead of an interference between system nonlinearity and applied time step, they are present in many methods, including the well-known Stoll \& Schneider \cite{SS}, Brooks \& Br\"unger \& Karplus \cite{BBK}, and Pastor \& Brooks \& Szabo \cite{Pastor_88} methods. However, given that the GJF framework offers the opportunity to conduct statistical simulations without the usual systematic increase in time-step errors, the effects of the resonant modes on the statistics become more pronounced, as seen in Figs.~\ref{fig_1} and \ref{fig_2}.  We also reemphasize that simulations with larger normalized damping values will depress both stability and existence of the ghost modes, thereby return the statistical sampling in discrete time to the expected physical results.

We finally comment that the results and considerations in this paper have been exclusively generated in light of stochastic thermostats without kinetic energy feed-back, such as what is done in, e.g., the deterministic Nos{\' e}-Hoover thermostat \cite{Nose,Hoover2,Holian}. One can expect that the feed-back of excess kinetic energy, which the modes in this paper possess, will be depressed by the algorithmic features, thereby suppressing the non-physical energetic ghost modes. It is, however, not obvious that this suppression will happen in accordance with physics, and we have here focused exclusively on recent stochastic thermostats that have been demonstrated to respond linearly correct to thermal effects.

\section{Acknowledgment}
The authors are grateful to Alan R.~Bishop and William G.~Hoover for their thoughtful comments to this work and its presentation.


\begin{thebibliography}{99}
\bibitem{comment_on_1000} The 1,000 independent simulations for each data point are necessitated by the calculation of the diffusion constant, since a single trajectory is a poor representative of diffusion. The obtained statistics is well represented by a single trajectory, and there is no statistical significance between simulations of different seeds for the random number generator, which is different for each simulated trajectory.
\bibitem{sqrt_b} Notice that the definition of the statistically sound half-step velocity $u^{n+\frac{1}{2}}$ Eq.~(\ref{eq:2GJ_r}) is slightly scaled such that a ballistic trajectory will be measured by $u$ as enhanced by $1/\sqrt{b}$.
\end{thebibliography}
\end{document}